# Are e-readers suitable tools for scholarly work?

**Siegfried Schomisch[1], Maria Zens, Philipp Mayr**

## Abstract

**Purpose** – This paper aims to offer insights into the usability, acceptance and limitations of e-readers with regard to the specific requirements of scholarly text work. To fit into the academic workflow non-linear reading, bookmarking, commenting, extracting text or the integration of non-textual elements must be supported.

**Design/methodology/approach** – A group of social science students were questioned about their experiences with electronic publications for study purposes. This same group executed several text-related tasks with the digitized material presented to them in two different file formats on four different e-readers. Their performances were subsequently evaluated by means of frequency analyses in detail.

**Findings** – e-Publications have made advances in the academic world; however e-readers do not yet fit seamlessly into the established chain of scholarly text-processing focusing on how readers use material during and after reading. Our tests revealed major deficiencies in these techniques.

**Research limitations** – With a small number of participants (n=26) qualitative insights can be obtained, not representative results. Further testing with participants from various disciplines and of varying academic status is required to arrive at more broadly applicable results.

**Practical implications** – Our test results help to optimize file conversion routines for scholarly texts. We evaluated our data on the basis of descriptive statistics and abstained from any statistical significance test.

**Originality/value** – The usability test of e-readers in a scientific context aligns with both studies on the prevalence of e-books in the sciences and technical test reports of portable reading devices. Still, it takes a distinctive angle in focusing on the characteristics and procedures of textual work in the social sciences and measures the usability of e-readers and file-features against these standards.

### Introduction

E-books are on the rise in science and libraries and are becoming serious competitors to the classic printed book. The literature offers different definitions of e-books extending from an electronic monograph up to “any piece of electronic text, regardless of size or composition (a digital object), but excluding journal publications, made available electronically (or optically) for any device (handheld or desk-bound) that includes a screen" (Armstrong, Edwards et al. 2002).

[1] Corresponding author: siegfried.schomisch@gesis.org

JISC generally applies the term “e-book” “to generic e-books available via the library, retail channels or on the web” (JISC, 2009, p. b). The following definition is more detailed and differentiated: “(1) An e-book is a digital object with textual and/or other content, which arises as a result of integrating the familiar concept of a book with features that can be provided in an electronic environment. (2) E-books typically have in-use features such as search and cross reference functions, hypertext links, bookmarks, annotations, highlights, multimedia objects and interactive tools.” (Vassiliou and Rowley, 2008, p. 363).

E-books represent a radical change in reading and usage of scientific texts. Typically, scientific texts possess high information density. For text understanding, selective techniques are essential for successful study. “Read only” appears insufficient in a scholarly context, additional features for printing, marking, annotating, and excerpting are crucial for textual work in academia.

Digital libraries have advanced and promoted the digitization of books (Candela et al., 2011), storing texts in different digital formats and providing a search interface to locate these digital resources. This development was accompanied by the rollout of portable e-book readers beginning in 1999 with the “Rocket eBook” in the U.S. market. The introduction of the Amazon Kindle in 2007 was the decisive breakthrough for e-readers and was soon followed by others (Sony, Barnes and Noble etc.). The introduction of the Apple iPad in 2010 as a tablet portable device was the next important innovation for using digital e-books online.

Many market analysis comparisons, both general and international, of surveys (e.g. Primary Research Group, 2009 and 2011 and NACS OnCampus Research, 2011) and overviews (e.g., the SWOT analysis of e-books and e-readers in Moss, 2010, p. 58) examined consumer attitudes towards, and interest in digital reading (Grzeschik et al., 2011; Mussinelli, 2010; PricewaterhouseCoopers, 2010 and Vassiliou and Rowley, 2008). The number of adults who own an e-reader increased in the United States from 6% in November 2010 to 12% in May 2011 while tablet computer ownership grew more slowly in the corresponding period (Purcell, 2011).

Our survey and user test was an attempt to gain insights into the usage and acceptance of e-books in typical scientific reading scenarios (see also Koch et al., 2010). The main issue is whether any of the commonly selected e-readers and/or the iPad can be used for typical, simple scientific work with literature. We examined the usability, handling and performance of these devices based on practical tests. Our results are discussed in the context of similar studies.

## State of the art

In the literature a number of scientific approaches towards e-book research emerge from information and library science and other cross-disciplinary academic areas, like psychology or pedagogics. These studies deal with the use (e.g. Shin, 2011) or design (e.g. Chong et al., 2009) of e-books in comparison to printed texts (e.g. Van der Velde and Ernst, 2009; Shelburne, 2009) and also with e-reader usability/ adoption (e.g. Gibson and Forbes, 2011; Gielen, 2010; Gupta and Gullett-Scaggs, 2010; Lai and Chang, 2011; Lam et al., 2009; Pearson et al., 2010).

Patterson et al. (2010) evaluated the particular requirements of learning and literacy in an innovative Canadian school during the development and implementation of a customized online library for "digital native" students. This research project combined information literacy issues, e-pedagogy and computational modeling activities and summarized the pros and cons of the e-readers and online library having an impact on learning and teaching. In another study on distance and work-based learning 28 Sony PRS-505™ e-book readers were pre-loaded with course materials and sent out to students. These devices yielded enhanced flexibility in curriculum delivery, improved efficiency in the use of study time, new strategies for reading course materials and cost reductions (Nie et al., 2011).

An online survey in the UK education and library sector (EDUSERV, 2010) asked mainly persons with a higher education about current and future institutional attitudes concerning e-books. The study emphasized e-book budgets, purchasing and associated decision making. Even though current usage is relatively low, the respondents predicted a higher rate of growth for e-books over the next two years (see also Mumenthaler, 2010).

A task-oriented survey (Berg et al., 2010) based on a think-aloud method revealed the different experiences and information retrieval behavior of Canadian undergraduate students in using e-books compared to print texts. Contrary to expectations the computer literate probands did not navigate and use e-books more intuitively and effectively; and they "were unclear about both the structure and functionality of e-books" (Berg et al., 2010, p. 524). The students searched information in print books in a more direct manner than in e-books and many were dissatisfied with the search function in e-books, because it did not work like a Google search.

Another project (see JISC, 2011) evaluated the potential of e-readers, which are designed for reading books, to replace paper copies in committee meetings. A Sony Touch e-reader was issued to all probands along with a limited number of Apple iPads. Most committee members used a VDU (= Visual Display Unit) screen in their preparation. Compared to the iPad, the Sony E-Reader was not a suitable device for committee preparation because of the small screen size, slow page turning and its inability to legibly display some formats. The probands preferred to choose their familiar devices, including laptops, netbooks, tablet PCs etc.

A university level pilot program in a classroom setting (Princeton University, 2010) investigated whether using Amazon Kindle DX e-readers could reduce the amount of printing and photocopying by users. E-reader users printed approximately half the amount of sheets compared with control groups not using the device. The Kindle DX "reading" experiences were rated better than "writing" experiences, a result which suggests that future e-book manufacturers should pay more attention to annotation tools, pagination, content organization, and achieving a more natural “paper-like” user experience.

E-book design and function were evaluated by Landoni (2010) in an effort to improve their quality, while Armstrong and Lonsdale (2009) mainly analyzed the attitudes to and preferences for using and reading print vs. electronic textbooks in academic work. In contrast to printed textbooks, e-textbooks were not used for extended reading but did receive high marks for their interactive features and facility when searching for information. Regarding social networks, Martin and Quan-Haase (2011) particularly emphasize the role of librarians as agents of change in the process of historians adopting e-books. Historians showed both positive (eagerness and curiosity) and negative (degree of reluctance and skepticism) attitudes towards adopting certain e-book features.

A current study (Shen, 2010) of student habits and attitudes found a growing readiness to access e-books via handheld devices and an increasing use of e-books. Professionals and students at Helsinki University of Technology Library tested various e-readers to see what demands and restrictions were being placed on e-materials, and how well current e-collections of the library can be used on e-readers (Aaltonen et al., 2011).

From the users’ perspective, immediate and ubiquitous access is already a strong argument in favor of e-books, an aspect reinforced by lightweight portable devices now enabling “mobile scientific working.” There appears to be no question that e-books “have entered into the mainstream of academic life and people are increasingly expecting to source e-book materials from their university library” (JISC 2009, p. 13).

Our study focuses on current user attitudes towards the use of e-books and e-readers in a typical scholarly context.

## User study

We expect to obtain qualitative results on the usage and acceptance of e-books and e-readers with regard to the specific patterns of information reception and processing in the social sciences. Through a *qualitative* investigation we would like to complement the aggregated survey results on the ever increasing use of digital content which concentrates on, for example, the *quantitative* amount of borrowing or selling. The usability of e-readers in our study is being evaluated on the grounds of task-based tests which encompass some specialties in the handling of scientific texts. Results from a qualitative usability test provide evidence to the extent that typical e-books from the social sciences - presented as EPUB and PDF-files on various portable reading devices - fit into the academic workflow. This study highlights both the added value and the shortcomings of electronic books compared with the classic “paper-based” book. An attempt is also made to identify desirable features of both e-readers and file formats that would enhance their suitability for students and scholars.

The reception of scholarly texts differs from everyday reading; whether for information or leisure. E-readers must meet a variety of functional requirements in different reading contexts. For instance reading a novel is a *consumptive* activity, so the device must mainly support direct linear readability. Scholarly text work, in contrast, is *productive* in the sense that the text is not only read but used to produce something new: a new text, new knowledge etc. For the needs of scholars and students text-related techniques go beyond linear readability as typical patterns include selective reading, non-linear reception, quick relevance assessing, re-reading, annotating, extracting, archiving, and further processing of the material in their own writings. These are well established workflows and generally taken for granted in the paper-based world, but how do they translate into that of e-books and the current mobile devices? Specific elements like annotations, tables, figures or formulas present particular challenges to the layout.

Our user test investigates whether e-readers can be used as an alternative or a complementary device for scientific reading. We seek to obtain more information on how texts in social sciences should be digitally prepared and converted.

Central to the idea of e-books is the choice of the right file format in the right place: Does the advantage of a “fixed-page view,” in the case of PDF, which is highly appreciated - especially in the layout and print area - produce usability problems in the mobile device display? Is the EPUB format the better solution, because “EPUB allows digital publication text to reflow according to screen size, enabling the publisher to distribute and the reader to consume digital publications on a variety of screen sizes” (Adobe, 2008)? Or is the “right format” a matter of adequacy and depends on the kind of content to be displayed? To tackle this issue we asked the participants (see next section below) to assess the suitability of both PDF- and EPUB-formats applied to different types of (GESIS) publications.

Our questionnaire and the subsequent testing tasks aim to address the terms of text reception in social science as a matter of both scholarly practice and technical requirements. Therefore we hypothesize:

*H1*. Scholarly reading procedures differ from everyday reading. Non-linear reading and extended options for the processing of text elements are crucial. The suitability of e-readers for scholarly work depends on their capacity to support these practices.

Many empirical studies, such as those of library and information science students (Pattuelli and Rabina, 2010) have been conducted with the objective of evaluating and comparing the usability of different e-readers (see e.g. Gibson and Forbes, 2011, Lai and Chang, 2011). In our study we would like to show the extent to which these field-specific requirements and the test outcomes for these tasks determine the overall evaluation of e-readers as suitable tools in an academic workflow. If the “reading plus” elements named above are crucial, their assessment should outperform the more general functionalities of solely the task of reading.

Taking these task- and workflow-based assumptions further into the realms of hard- and software performance, one might presume that the more flexible EPUB-format and the more adaptable tablet computer perform better than their respective competitors. Our hypotheses are:

*H2*. In a user test we expect the genuine e-reader EPUB-file format to be more suitable for scholarly requirements than the PDF-format.

*H3.* We assume that the tablet computer will obtain better user approval and acceptance than the dedicated e-readers.

The role and the influence of gender and related stereotypes in the processes of perception, acceptance and adoption of new technologies and information systems have been examined and discussed in a multitude of studies. The results are mixed, but overall it could be argued that in the academic world, gender-related effects in ICT use appear to be less significant than in society as a whole, and confined more to self-efficacy than actual performance (e.g. Booth et al., 2010; Grohman and Battistella, 2011; Si and Man, 2010). Since our study had an even distribution of male and female participants we desired a closer look at their approaches and performances starting with this hypothesis:

*H4.* We do not expect gender-related differences in the usage and acceptance of e-readers and e-books or the completion of practical task-based tests among students.

**Test procedure**

The qualitative investigation comprised a written survey along with task-based tests completed by a small group of 26 persons. Both the questionnaire and the tasks were pretested and modifications were made regarding any content comprehension issues. The tests were performed using three dedicated e-readers: “Amazon Kindle 2” (Software Version 2.3), “Sony Reader PRS-600 Touch Edition” (Firmware Version 1.3), and the “Onyx Boox 60 with WLAN connection“ (eBook Reader Library Version 3.2). The fourth device was an Apple TabletPC iPad (Version 3.2).

To obtain the most homogenous test group possible we selected mostly social science students (12 female and 14 male). The majority of test persons (n=18) were somewhere between their third and tenth semesters in their studies, with the average number of semesters being 8.4. Only one person had previous experience with e-readers prior to the test.

The students were divided into four test groups of six persons each; one group for each device. Three of these groups were tested in the cognitive lab of the GESIS

Survey Design and Methodology department (pretesting area) located in Mannheim/Germany. The tests were recorded on camera. There was a technical fault with one Amazon Kindle2, so only two tests of this device were successfully finished and included in the overall evaluation. The other groups were then increased to include eight persons. The fourth group was tested in the University Library of Cologne using an iPad TabletPC. These tests were not recorded on camera.

The probands were individually surveyed using the questionnaire. All groups demonstrated a very high willingness to respond. A device-specific user guide was available on a notebook for the probands. There was a test supervisor present at all times to assist with any technical questions. These persons subsequently conducted conversations with each test person. Video recordings were also used for the survey evaluation. The individual tests lasted 40-50 minutes on average, with small device-specific differences.

Each proband received an e-reader containing all the test documents, in addition to the questionnaire; specifically, these included three social science texts and two journal articles.

Both of the included journal articles [see I and II] contain several figures, tables and formulas, slightly larger than one A4 format page, as well as a two-column layout in one chapter. The third test document [see III] is a bibliography with six chapters where references and descriptions of research projects on a particular issue are arranged. The content is mainly continuous text but also contains several indices with jump marks to the corresponding text passage as well as hyperlinks in the text. The source files of the texts to be evaluated – Adobe Indesign (*.indd) for the journal articles and OpenOffice Writer (*.odt) for the information service, respectively – were exported into the EPUB format using the export features of the application software mentioned while also installing the writer2xhtml-Extension in OpenOffice Writer.

We tried to verify our hypotheses in the context of frequency distributions. In consequence of the small sample size (n=26) we evaluated our data on the basis of descriptive statistics and abstained from any statistical significance test.

**Results**

While highlighting the degree of utilization of e-books and e-readers, we attempted to determine if the functionalities and usability of e-readers suffice for scholarly reading. The pros and cons of the EPUB file format were compared against PDF and the performance of the iPad against dedicated e-readers. Gender-related differences in e-reader usage and acceptance were also addressed.

*User behavior*

The majority (n=22) of the probands indicated that they do not search selectively for e-books. However 15 of the respondents said that they have used e-books in the last six months. Research literature (n=12) tops the list (see Figure 1) of the (very) frequently utilized digital literature in this period, followed by text books (n=6) and reference or advisory books (n=4).

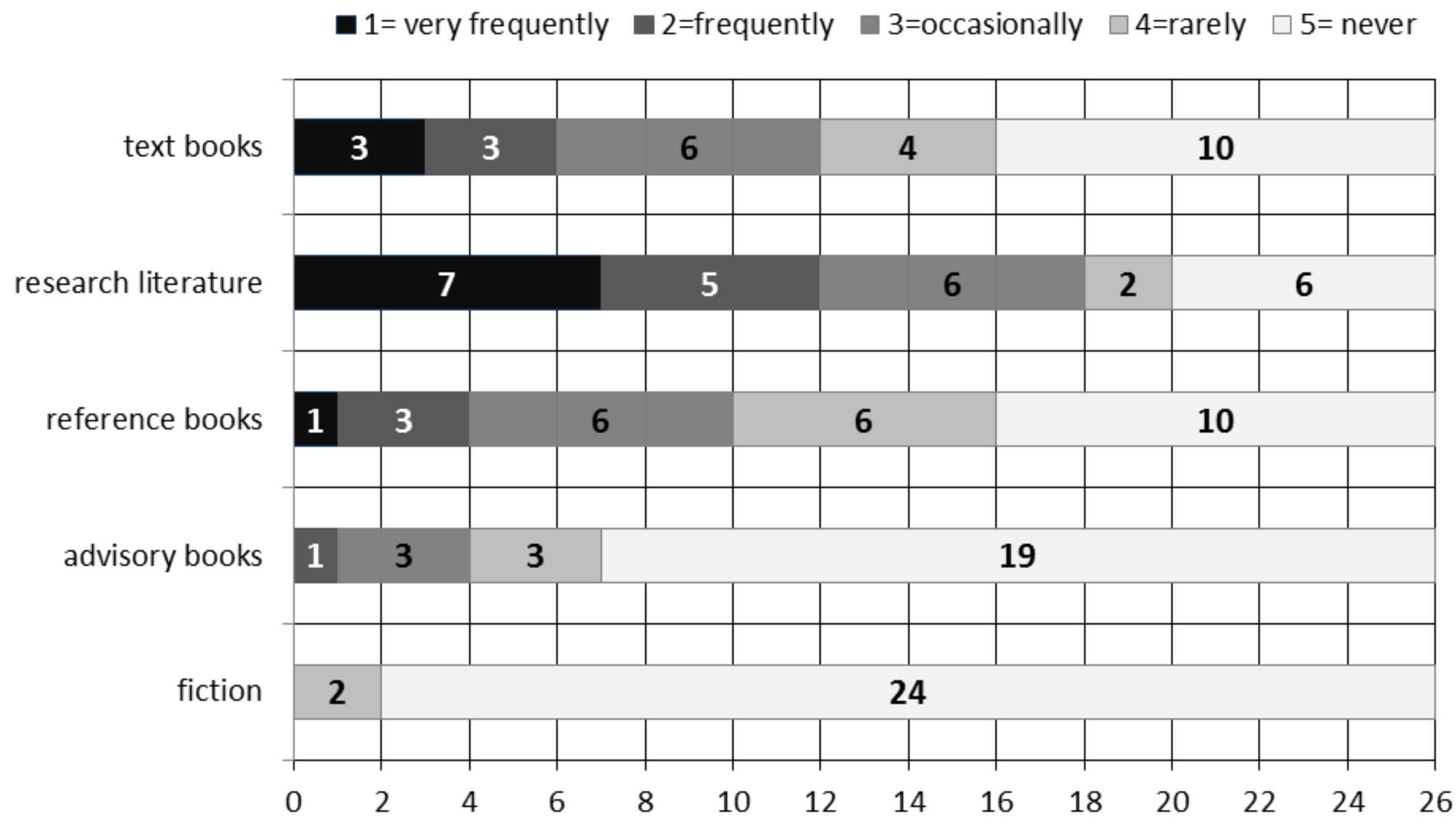

**Figure 1: Frequencies of reading e-books in the last six months**

In terms of locale, the majority (n=21) read primarily at home followed by the library (n=14). Only three probands indicated that they also read e-books "on the go" (such as on vacation, in cafés, or during their commute, for example etc.). These persons belong to the small group also utilizing mobile end devices like smartphones as a reading device (n=4). Apart from that, laptop and PC are the most often cited reading devices used (see Figure 2).

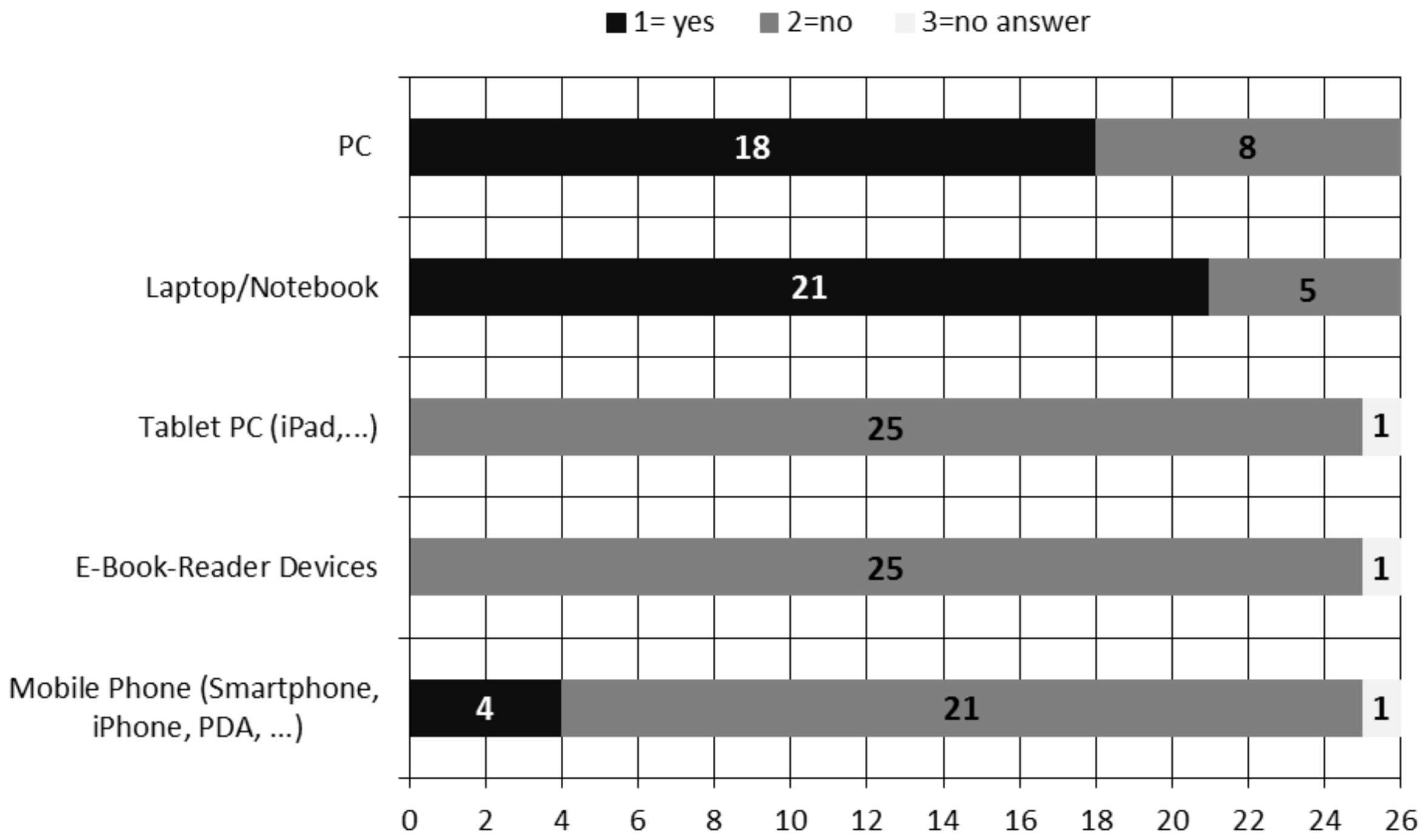

**Figure 2: Devices used for reading e-books**

The test persons' reading habits are predominated by "power-browsing"; the rapid looking up of information (n=21). 18 out of the group stated that they do not use e-books for intensive onscreen reading, indicating that important parts of texts were printed (n=18).

*Functionalities of the e-reader and usability*

Supporting active examination of the text during the process of reading belongs to the basic functionalities of all devices tested. However device-specific differences were found in the implementation of certain functions, as well in dependence to the file formats supported. For example, Amazon's Kindle2 demonstrated difficulty with highlighting and annotation in PDF while the Onyx Boox 60 lacks an annotation function for the EPUB-format.

Following the test exercises working with particular e-reader functions, probands were then asked to assess the usability of the technical implementation (black bar in Fig. 3) and to also rate the importance of these functions for them while working with e-books on a scale with the five functionalities tested (grey bar in Fig. 3). This was done for both subjects indicating from 1= very user-friendly up to 5= not at all user-friendly or 1= very important up to 5= not at all important, respectively. Figure 3 shows the frequency distribution of the two questions.

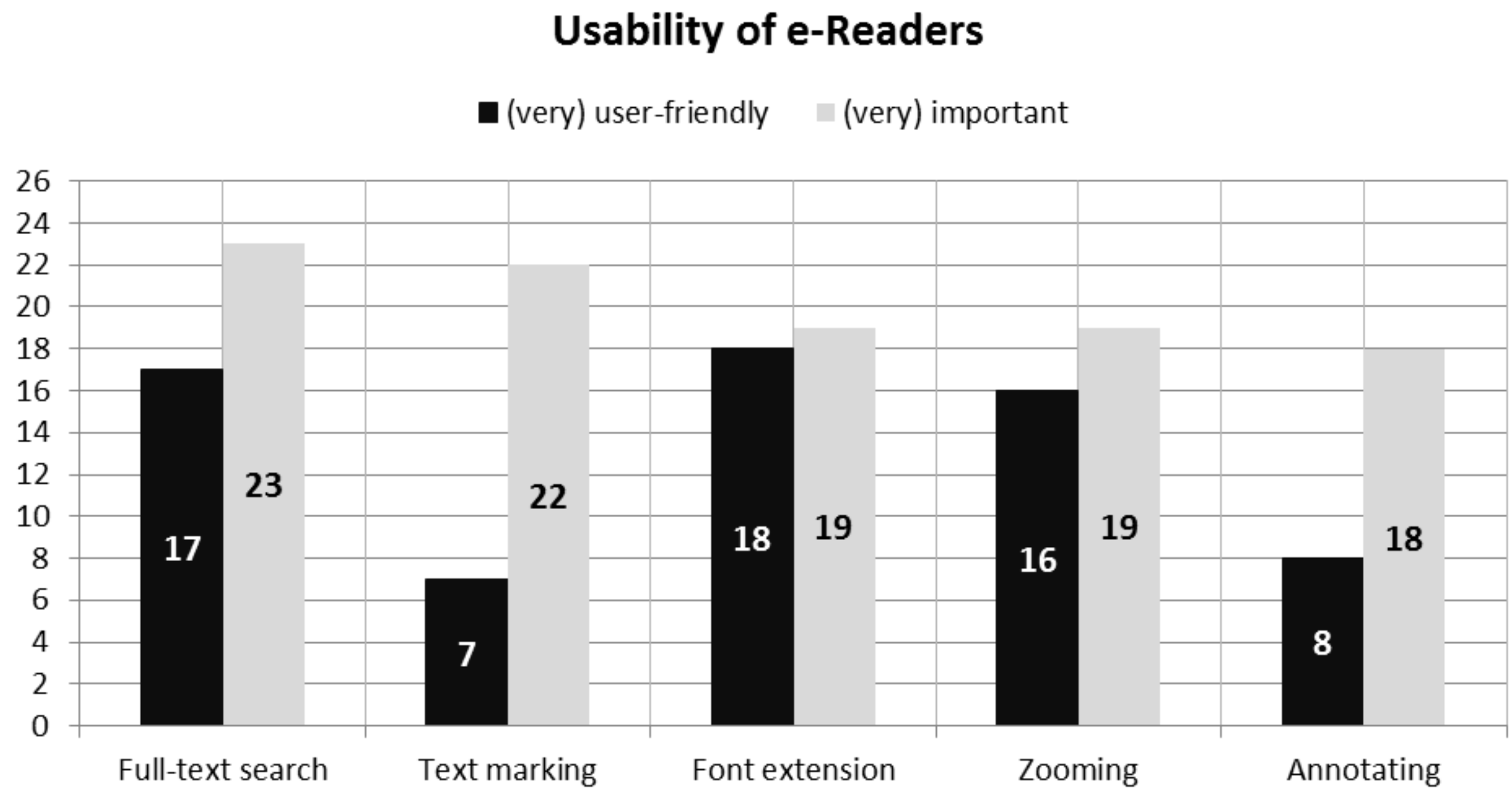

**Figure 3: Tested functionalities of e-readers**

A large discrepancy between usability and importance in text marking and annotating is immediately apparent. Font extension, full text search and zooming displayed the highest positive correlation in terms of usability. In terms of importance, full text search leads, closely followed by text marking. This is not surprising because search functionalities are a substantial added value of e-books compared with printed books. Nevertheless, the probands indicated marginal shortcomings in searching with the e-readers, potentially indicating usability problems.

This applies particularly to the poor performance in text marking and annotating and was confirmed by replies such as those from the Onyx test group: “You can’t actually write really legibly with the pen. Word recognition software would be a great help.” One proband of the iPad-group felt that the “possibility of marking more than one page” was lacking.

We were also interested to learn how important the printing and exporting of text was for the test subjects. The answers were that both were (very) important: for 22 respondents “printing text” and for 20 respondents “exporting text”.

Another battery of questions addressed functionalities related to web2.0: user sharing (exchange of annotations via Internet); e-mail in order to recommend a book to friends; Internet access to go to hyperlinks in the text. User sharing and e-mail were classified by 21 and 17 probands, respectively, as (generally) not important, whereas 20 respondents, a clear majority, mentioned the direct connection to hyperlinks from the text as (very) important.

Our findings regarding H1 indicate that the functionalities and usability of the e-readers tested are neither sufficient nor suitable for the demands of scholarly work.

*PDF versus EPUB*

To compare the two different file formats probands were asked to open a predetermined text on the given test device. Text navigation was tested via selective tasks (searching a text passage, jumping to/from parts of a page, directing to figures, formulas and tables). These working tasks were carried out with font size adjustment of the devices first set to default and then utilizing the font extension and zoom function. Evaluations were recorded following each task using the prescribed rating scale (see Figures 4 and 5). The Kindle test group was excluded from this comparison as the Kindle e-reader does not support the EPUB format

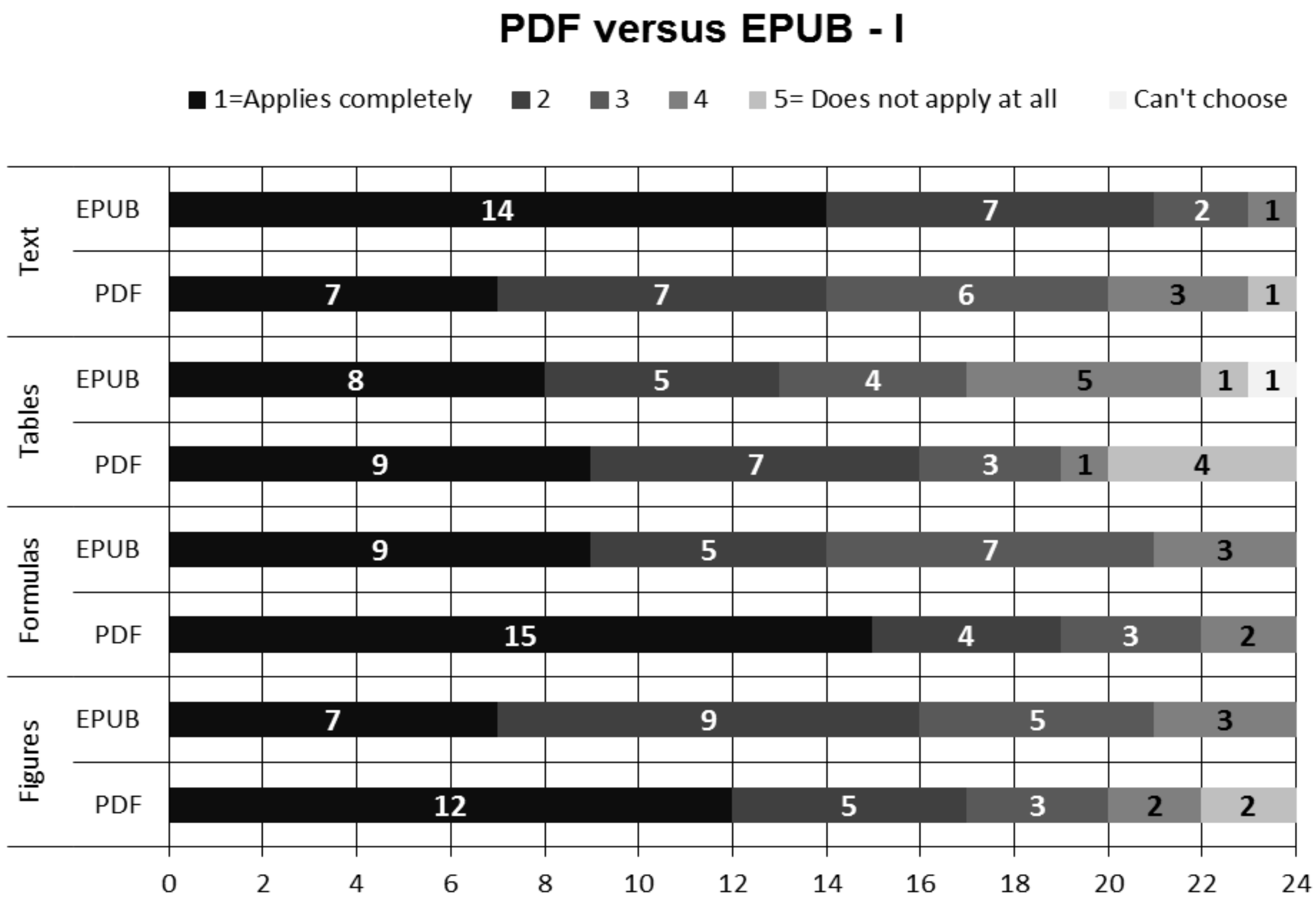

**Figure 4: Comparison of the file formats using default setting of the devices**

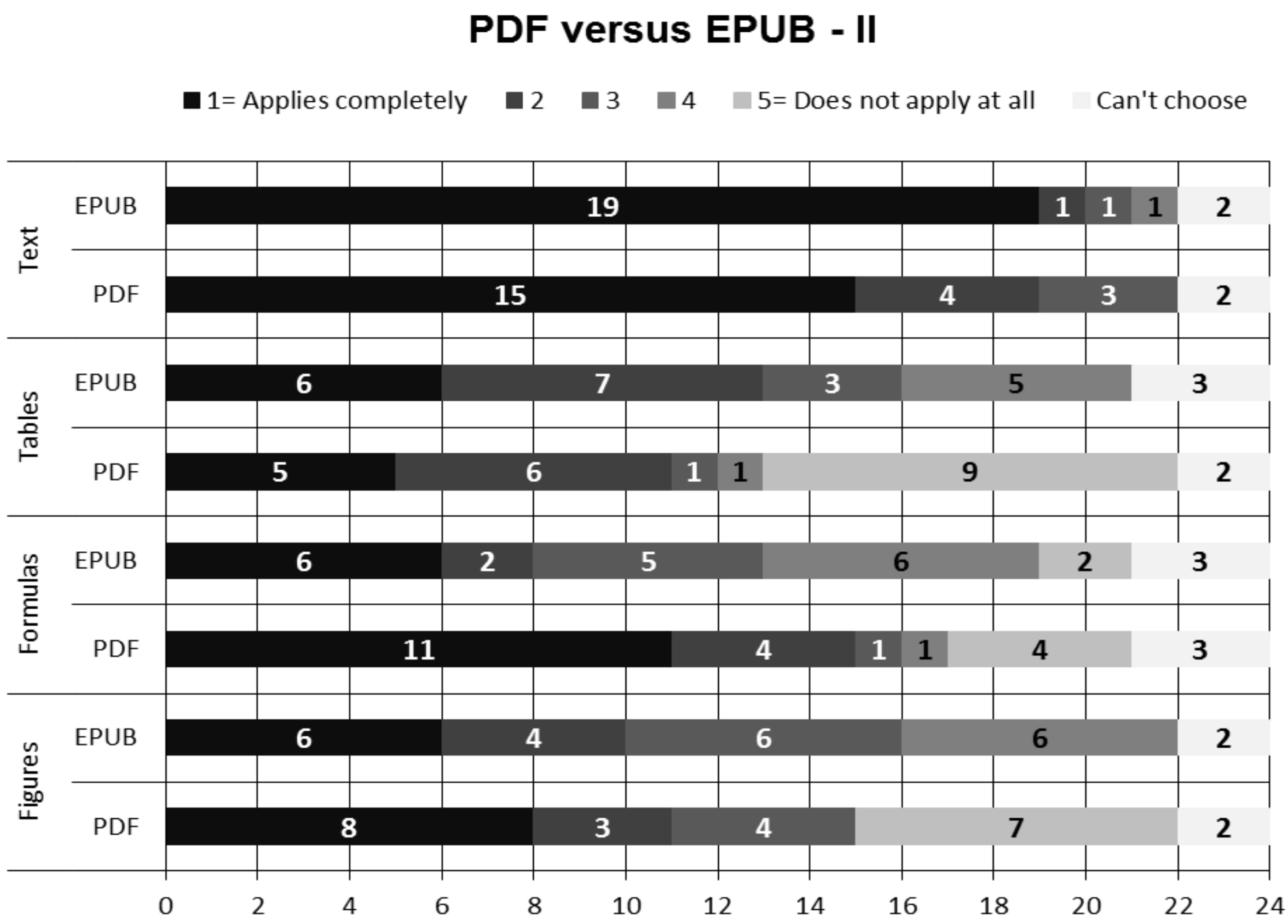

**Figure 5: Comparison of the file formats using font extension or zooming**

The EPUB format clearly outperformed the others when working with text. A majority of users indicated they found the text to be (very) legible and easy to read both in the default font size (Figure 4) and in extension and zoom modes, respectively (Figure 5)..

There was greater differentiation in statements given in response to questions on the display of figures, formulas and tables; a possible point for further analysis.

Our finding here is that the results on this topic corroborate our second hypothesis (H2) assuming EPUB would be the favored file format only in terms of readability of textual information.

*Color and Display*

Assuming that colored displays of figures and graphics are generally desired in scientific texts, it came as a surprise to find that 10 of 26 respondents were at least indifferent towards a colored display. We might see a different result if the probands were selected from natural sciences or technical disciplines, which utilize more figures etc. and less textual elements. However, all probands definitively emphasized the desirability of an eye-friendly display (n=23: very important; n=3: important).

*Are e-readers an alternative to reading on a PC?*

We asked: “Is the examined e-book-reader an alternative for you or a reasonable supplement to reading on a PC with regard to reading scientific texts as utilized in the test?” Of the 26 respondents, 14 replied no and 12 yes to this question.

The following results (differentiated by device) were obtained: the majority (n=11) of 18 probands testing the dedicated e-readers Kindle, Sony and Onyx stated that the e-reader was not an alternative for them, while seven probands indicated yes the device was an alternative. Further differences were noted by device. Thus the result in the Sony group was clearly undecided, while in the Onyx group, five of a total of eight testers responded no, the device was not an alternative. In the iPad group five of a total of eight probands indicated yes, and three no. Here, neither concordant refusal nor acceptance was stated. The different reading experiences of the probands during the test decidedly influenced the final test results. This includes the practical experiences with “frozen” devices (Sony once, Onyx twice) where a reset was required before continuing, and the total crash of the Kindle.

Responses given indicated that the expectations for e-readers were strongly determined by previous PC experience:

“An interface like windows is missing so that you also have a browser etc. at your disposal; faster scrolling … cutting of particular sections … (desired); zooming over the menu is laborious.”

Our hypothesis (*H3*) assuming the superiority of the iPad (software version 3.2) over the dedicated e-readers in both practical and task based tests was disproven by our results.

The users of the iPad evaluated the functionalities “font extension” and “zooming,” for example, equally to those found on the Sony and Onyx devices with respect to their technical implementation and user-friendly handling. The results for these e-readers were similar with regard to estimating the appearance of figures in PDF format with the default settings of the devices, and also for evaluating whether the text is easily readable by utilizing the font extension and zooming in both the PDF and EPUB formats. iPad users evaluated the functionality “writing annotations” on their devices far better than did users of other devices with 87,5 % of iPad users rating this (very) user-friendly in contrast to only 4% of the Sony users (the only ones in this range of the dedicated e-readers).

Our frequency analysis displayed no gender-related differences in terms of technical affinity, usage of e-books or adoption of the e-readers.

We can state that, in total, our test results showed no gender-related differences; as evinced by the video recordings.

## Conclusion

The goal of our user survey was to gain knowledge on readability and usability of e-readers in a scientific context. In consequence of the small sample size (n=26) we evaluated our data on the basis of descriptive statistics and abstained from any statistical significance test. Essentially only one proband in our study had at least occasionally used an e-reader before starting the user test and most of the probands indicated having previously read e-books on Laptops/Notebooks and PCs at home or in the library. However, these results seem to change under other terms and conditions, as shown, for example, in the study of Nie et al. (2011) where students tested e-readers in distance work-based learning. These test persons tended to read the course materials on their e-readers more on the move (e.g. train, bus, plane) or in public places (e.g. library, café).

The considerable usability problems encountered with the e-readers tested calls into question the suitability of the dedicated e-reader for scientific reading and working. Gibson and Forbes (2011) obtained similar results in their “second-generation” e-reader evaluation despite technical improvements in size, weight and screen quality. The Sony Touch Edition (PRS-600) tested in the study by JISC (2011) lacked most of the functionality for reading PDF files in an acceptable format when using the magnify option. While the iPad received a better evaluation for its larger screen, page turning and text enlarging functionalities, etc. The theory-driven evaluation of Lai and Chang (2011) underscores the relevance of convenience, compatibility and media richness for the adoption of dedicated e-readers.

Significantly, we found that the appearance of the texts on the e-reader displays, their readability and legibility, indicated the strengths and weaknesses in the PDF and EPUB formats. Conversion methods in particular should therefore be optimized for scientific texts when creating appropriate formats for mobile devices to minimize display problems independent of the devices (see Gielen, 2010). This primarily concerns pictures, graphic illustrations, tables and formulas in the EPUB format. The question remains open whether e-readers based on, for example, HTML5-functionalities (see Georgiev et al. 2011) in combination with e-books in the new EPUB 3 format (see IDPF, 2011 and Williams, 2011) or in the currently announced Kindle Format 8 [see IV] will prospectively advance the usage and acceptance of e-books and e-readers among the scientific community. It may also be worth noting whether the flexibility of the EPUB format is better suited to e-reader devices, and thus, will find increased reception among readers. What can’t be overlooked, however, is that the PDF text format, because of its fixed nature, is the widely accepted long-term digital archiving format for repositories (Seadle, 2009).

Important basic functions for working with text such as marking and annotating were evaluated as not very user-friendly in e-reader tests. The Onyx device tested lacked any further processing features such as exporting marked paragraphs and annotations. These functionalities were rated as insufficiently implemented in both of the other e-readers (Kindle and Sony) evaluated. Lam (2009) confirms these results especially for pocket personal computers (PPCs) tested, as was similarly found by Nie et al. (2011) in tests of the Sony PRS-505™. Patterson et al. (2010) also criticized the unwieldy and difficult annotation tool. The majority of our probands evaluated having a print function as (very) important. This finding corroborates their reading behavior with their answers (important parts were printed; on the screen, there is no intensive reading). In contrast, Nie et al. (2011) stressed in their evaluation that some e-reader probands became less dependent on printed materials, or more selective in what they eventually printed.

No significant gender-related differences were detected with regard to acceptance or usage of e-books and e-readers. In her e-book user evaluation integrating emotional (affective) and cognitive factors, Shin (2011) noticed that gender played no significant role in the adoption or usage of e-books.

Given that it received user evaluations similar to the Sony and Onyx, the iPad did not verify our hypothesis H 3 for it to be the more accepted and user-friendly device when compared with dedicated e-readers; in contrast to findings by JISC (2011) and also Aaltonen et al. (2011).

The results of our user tests confirm the assumption that the dedicated e-readers could not really compete with all-purpose devices like PCs, tablets or notebooks (see Mumenthaler, 2010 and Aaltonen et al., 2011). The dedicated e-readers we used in our tests were found to be unsuited for the special reading techniques and demands of scientific work.

**Future research**

Apart from the detailed critique of tested features and performances, we observed a great willingness by our test subjects to try and add new devices to their academic toolbox. Our study – given its limitations and the small sample size - shows that there is a basic acceptance and positive attitude to use e-readers for scientific work. However, further technical improvement, more user-friendly equipment – possibly the interactive features of Apple iBooks 2 [see V] - and additional empirical research seems necessary. The latter might differentiate, for example, between scientific disciplines and/or user-dependent attitudes, behaviors, working environments, etc.

The progress of technical possibilities of devices (see Dougherty, 2010) is also associated with the question of the further development of e-books themselves, for example, with the implementation of cloud based e-book relevant services (see Shen and Koch, 2011) or digital scholarly communication.

Perhaps current neuroscience based results [see VI] might provide new cognitions about reading processes on reading devices versus paper concurrently measured by eye tracking and electrophysiological brain activity.

**Notes:**

[I]
http://www.gesis.org/fileadmin/upload/forschung/publikationen/zeitschriften/mda/Vol.3_Heft_2/07_Koch_et_al.pdf (accessed 08 December 2011).

[II]
http://www.gesis.org/fileadmin/upload/forschung/publikationen/zeitschriften/mda/Vol.3_Heft_2/05_Ostermann_Luedtke.pdf (accessed 08 December 2011).